\documentclass[aps, preprint, groupedaddress, floatfix]{revtex4}

\usepackage{epsfig}
\usepackage{graphicx}
\usepackage{amsmath}

\begin{document}

\title{Modifying the Electronic Orbitals of Nickelate
  Heterostructures Via Structural Distortions}

\author{Hanghui Chen$^{1,2,3}$, Divine P. Kumah$^{3}$, Ankit S. Disa$^{3}$,
Frederick J. Walker$^{3}$, Charles H. Ahn$^{3,4}$ and Sohrab
Ismail-Beigi$^{3}$}

\affiliation{ $^1$Department of Physics, Columbia University, New
  York, NY, 10027, USA\\ $^2$Department of Applied Physics and Applied
  Mathematics, Columbia University, New York, NY, 10027,
  USA\\ $^3$Department of Applied Physics, Yale University, New Haven,
  CT, 06520, USA\\$^4$ Department of Mechanical Engineering and Materials Science,
  Yale University, New Haven,
  CT, 06520, USA}
   \date{\today}

\begin{abstract}
We describe a general materials design approach that produces
large orbital energy splittings (orbital polarization) in
nickelate heterostructures, creating a two-dimensional single-band
electronic surface at the Fermi energy.  The resulting
electronic structure mimics that of the high temperature cuprate
superconductors. The two key ingredients are: (i) the construction
of atomic-scale distortions about the Ni site via charge transfer
and internal electric fields, and (ii) the use of three component
(tri-component) superlattices to break inversion symmetry.  We use
{\it ab initio} calculations to implement the approach, with
experimental verification of the critical structural motif that
enables the design to succeed.
\end{abstract}

\maketitle

\begin{figure}[t]
\includegraphics[angle=-90,width=16cm]{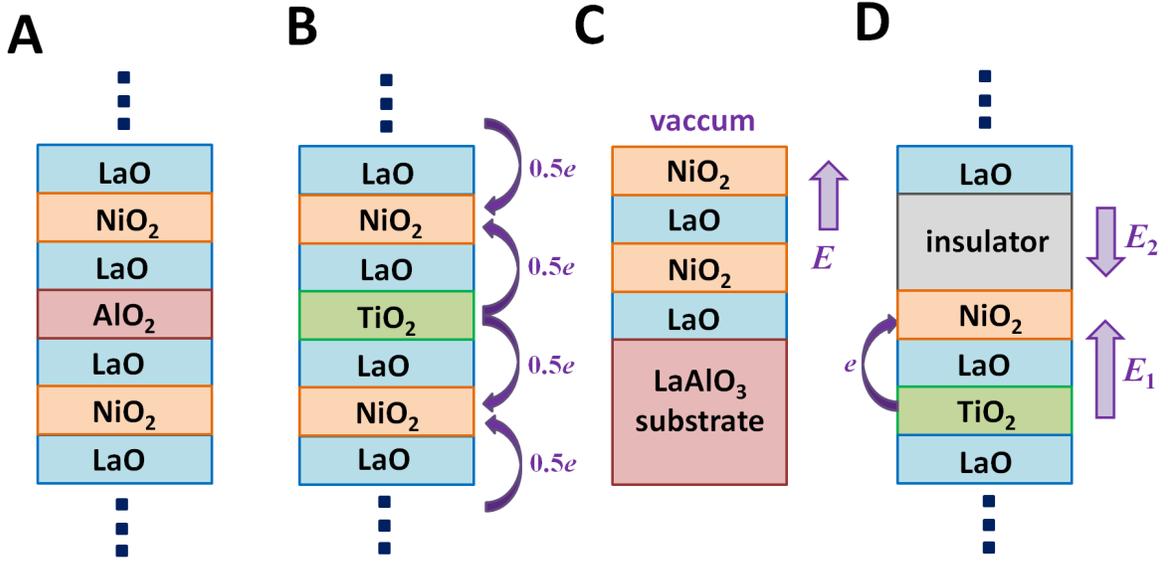}
\caption{\label{fig:schematic} Schematics of four nickelate
heterostructures. \textbf{A}) (LaAlO$_3$)$_1$/(LaNiO$_3$)$_1$
superlattice. \textbf{B})  (LaTiO$_3$)$_1$/(LaNiO$_3$)$_1$
superlattice with nominal electron transfer from Ti to Ni. \textbf{C})
NiO$_2$-terminated LaNiO$_3$ thin film on a LaAlO$_3$ substrate.
\textbf{D}) (LaTiO$_3$)$_1$/(LaNiO$_3$)$_1$/insulator superlattice
with electron transfer and broken inversion symmetry.  Arrows with
``$E$'' denote long-range electric fields in the material.}
\end{figure}

A forefront area in condensed matter physics involves the
modification of matter at the scale of individual atomic layers to
form artificial systems whose properties differ significantly from
their ``parent'' bulk forms.  Transition metal oxides are
paradigmatic: (i) their bulk forms display an array of physical
phenomena, including magnetism, metal-insulator transitions, and
superconductivity~\cite{wolfram_electronic_2006,
Zubko-ARCMP-2011}; (ii) in principle, one can choose among various
cations and their spatial ordering in the oxide lattice; and (iii)
in practice, advanced layer-by-layer fabrication techniques can
realize such
heterostrutures~\cite{ahn_electrostatic_2006,mannhart_two-dimensional_2008,
Rondinelli-AM-2011}. One current topic involves the engineering of
electronic states in heterostructures in order to emulate the
properties of high temperature copper oxide (cuprate)
superconductors.  A concrete proposal involves artificial
heterostructuring of rare-earth nickelate materials, specifically
atomically thin LaNiO$_3$ layers surrounded by insulating
LaAlO$_3$ layers in the (001) direction, to fulfill four basic
properties of carriers found in the cuprates and to realize a
single-band two-dimensional (2D) Hubbard model: spin one-half,
quasi 2D confinement, anti-ferromagnetic correlations, and lack of
orbital degeneracy~\cite{chaloupka_orbital_2008}.
Fig.~\ref{fig:schematic}A illustrates such a superlattice.  This
proposal has led to significant activity on nickelate
heterostructures~ \cite{Hansmann-PRL-2009, May-PRB-2009,
han_chemical_2010, Son_APL_2010, Scherwitzl-PRL-2011,
Chakhalian-PRL-2011, benckiser_orbital_2011, han_dynamical_2011,
freeland_orbital_2011, blanca-romero_confinement-induced_2011,
boris_dimensionality_2011, Han-PRB-2011}.

Experimentally, bulk LaNiO$_3$ is a metallic paramagnet with a
single electron in doubly degenerate $e_g$
bands~\cite{Sreedhar-PRB-1992, Eguchi_PRB_2009}. {\it Ab initio}
calculations confirm that two-component (bi-component)
LaNiO$_3$/LaAlO$_3$ heterostructuring reduces dimensionality by
reducing the band dispersion of the out-of-plane $d_{3z^2-r^2}$
band compared to the in-plane $d_{x^2-y^2}$
band~\cite{Hansmann-PRL-2009, han_chemical_2010,
benckiser_orbital_2011,han_dynamical_2011,freeland_orbital_2011,
blanca-romero_confinement-induced_2011}. The reduced dimensionality 
also enhances the correlation effects and a Mott transition 
is observed~\cite{Liu-PRB-2011}, in line with other oxide systems
where control over dimensionality and correlations can modify electronic
band structure~\cite{yoshimatsu-metalqwells-2011} or thermoelectric
properties~\cite{sparks-thermoelec-2012}.
Ultra-thin LaNiO$_3$
layers show a magnetic ground
state in both experiment and theory~\cite{boris_dimensionality_2011, 
Han-PRB-2012}. However, orbital
degeneracy is not much affected: pioneering
experiments~\cite{benckiser_orbital_2011,freeland_orbital_2011}
and {\it ab initio} calculations ~\cite{Hansmann-PRL-2009,
han_chemical_2010,blanca-romero_confinement-induced_2011} find
that the population of the two $e_g$ orbitals differ by
$\sim$5\%-10\%. Inclusion of Hubbard-type strong electronic
correlations on Ni can produce a significant difference of orbital
populations (i.e., orbital polarization) in a simplified effective
low-energy description~\cite{Hansmann-PRL-2009}, but when both Ni
and O orbitals are included in such a treatment, the orbital
polarization is significantly reduced~\cite{han_dynamical_2011}. 
In contrast, cuprates have 100\% orbital polarization which means that there are only
$d_{x^2-y^2}$ bands at the Fermi level, while the $d_{3z^2-r^2}$
bands are lower in energy and completely filled due to strong
crystal field splittings~\cite{cava_oxide_2000}. A key challenge
is to modify bi-component LaNiO$_3$/LaAlO$_3$ superlattices to
achieve similarly large orbital polarizations.

In this Letter, we describe a materials design approach that
engineers structural distortions in tri-component superlattices to
greatly enhance the orbital polarization. Fig.~\ref{fig:schematic}
shows the design schematics.  A first approach to overcoming the
small orbital polarization in LaNiO$_3$/LaAlO$_3$ superlattices is
to dope the LaNiO$_3$ layers: the $d_{3z^2-r^2}$ bands are narrow,
so filling (or emptying) them may move the Fermi level into the
$d_{x^2-y^2}$ bands.  Replacing the LaAlO$_3$ layers by LaTiO$_3$,
as displayed in Fig.~\ref{fig:schematic}B, can achieve this goal:
Ti$^{3+}$ in LaTiO$_3$ has one electron in its $d$ bands that
seeks the lower energy Ni sites and dopes them $n$-type. Below, we
describe that while the doping is effective, it is insufficient to
deliver full orbital polarization.  An alternate approach stems
from our and others' observations~\cite{Han-PRB-2011}
that the surface Ni atoms on NiO$_2$-terminated LaNiO$_3$ thin
films (Fig.~\ref{fig:schematic}C) have the desired large orbital
polarization due to eliminated bonds with the missing out-of-plane
(apical) oxygen. However, using such a surface system in practice
is challenging since the surface of a polar thin film can be
subject to various adsorbate perturbations that modify its
properties (cf.~\cite{wang_reversible_2009}). Nevertheless,
the main lesson of the thin film effect is to break or emulate the
breaking of a Ni-O bond, and to realize this possibility in a
three-dimensional superlattice, we design structural distortions
that greatly elongate the apical Ni-anion bonds sufficiently for
orbital engineering purposes.  We show that this approach is
realizable with a tri-component superlattice involving both the
LaTiO$_3$ doping and an added wide-gap insulator:  the
Ti$\rightarrow$Ni electron transfer, combined with the structural
asymmetry and the requirement of a periodic potential in a
superlattice, ensures the head-to-head electric field pattern
depicted in Fig.~\ref{fig:schematic}D. These fields move both
apical anions away from the Ni and create a strong orbital
polarization. In the remainder of this Letter, we use {\it ab
initio} calculations to describe each step of the process and use
experimental growth and characterization to verify the key
structural properties that deliver the large orbital energy
splitting.  We point out that {\it ab initio} calculations have successfully
described the large orbital polarization of high-temperature
cuprate superconductors~\cite{pickett_electronic_1989}.

Our theoretical work employs density functional
theory~\cite{Hohenberg-PR-1964,
  Kohn-PR-1965} within the supercell plane-wave pseudopotential
approach~\cite{Payne-RMP-1992}, the Quantum-ESPRESSO package
~\cite{QE}, the local density approximation
~\cite{Ceperley-PRL-1980, Perdew-PRB-1981}, and ultrasoft
pseudopotentials~\cite{Vanderbilt-PRB-1990}. The superlattice
direction and surface normal for thin films are along $z$. The
structures are periodic in the $x$ and $y$ directions. In most
of our calculations, the in-plane lattice
vectors are fixed to the theoretical one of LaAlO$_3$ at
$a=3.71$~\AA~(2\% smaller than experiment) in order to
simulate a LaAlO$_3$ substrate for superlattice
growth~\cite{strain}. All remaining structural degrees of freedom
are relaxed. We generate two $e_g$ maximally localized Wannier
functions ~\cite{marzari_maximally_1997,souza_maximally_2001} on
each Ni using the Wannier90 package \cite{wannier90}. By
construction, these functions reproduce the {\it ab initio}
anti-bonding conduction Ni $e_g$ bands and provide onsite energies
for the Ni $e_g$ orbitals. 
Due to the well known underestimation of
band gap in density functional theory, we use the rotationally
invariant LDA+$U$ approach~\cite{Lie-PRB-1995,Dudarev-PRB-1998},
with an accepted literature value of $U_{\textrm{Ti}}=4$ eV on
the Ti $d$ states~\cite{Mizokawa-PRB-1995}. The main purpose of
the $U_{\textrm{Ti}}$ is not to describe  correlated behavior
on the Ti (since it will turn out to be a fully ionized donor), but simply to
ameliorate the energy alignment between Ti and Ni $d$ states. A
detailed investigation shows a lack of dependence of the
main physical properties on $U_{\textrm{Ti}}$~\cite{bi-color}.
We purposely have not included
Hubbard $U$ corrections on Ni in our theoretical calculations since
LDA+$U$ calculations for nickelates lead to mixed results. For bulk
insulating rare-earth nickelates, $U_{\textrm{Ni}}
> 0 $ is necessary to yield an insulating and magnetic,
as opposed to a metallic and paramagnetic, ground
state~\cite{prosandeev_ab_2012,park_site-selective_2012}. However,
$U_{\textrm{Ni}} > 0$ for bulk LaNiO$_3$ worsens agreement
with experiment~\cite{gou_lattice_2011} (e.g., it predicts a
magnetic ground state, contrary to its paramagnetic
nature~\cite{blanca-romero_confinement-induced_2011, 
park_site-selective_2012,gou_lattice_2011}).
Important future directions involve seeing
how our predictions may be modified by $U_{\textrm{Ni}}>0$, 
for example by magnetic behavior on Ni, and, more importantly, to find
better ways of describing electronic correlations on the Ni site.
Experimentally, we grow four unit cell thick
films of LaNiO$_3$ on LaAlO$_3$ (001) substrates using molecular
beam epitaxy, with the structure determined by synchrotron x-ray
diffraction~\cite{supplement}.  The diffraction is analyzed using
the coherent Bragg rod analysis method~\cite{Yacoby-NatMat-2002}
to produce electron density maps. The atomic position for each
lattice site is identified by the centroid of the electron density
near each site in the perovskite lattice.
Further technical details on both theory and experiment are found
in~\cite{supplement}. 

\begin{figure}[t]
\includegraphics[angle=-90,width=14cm]{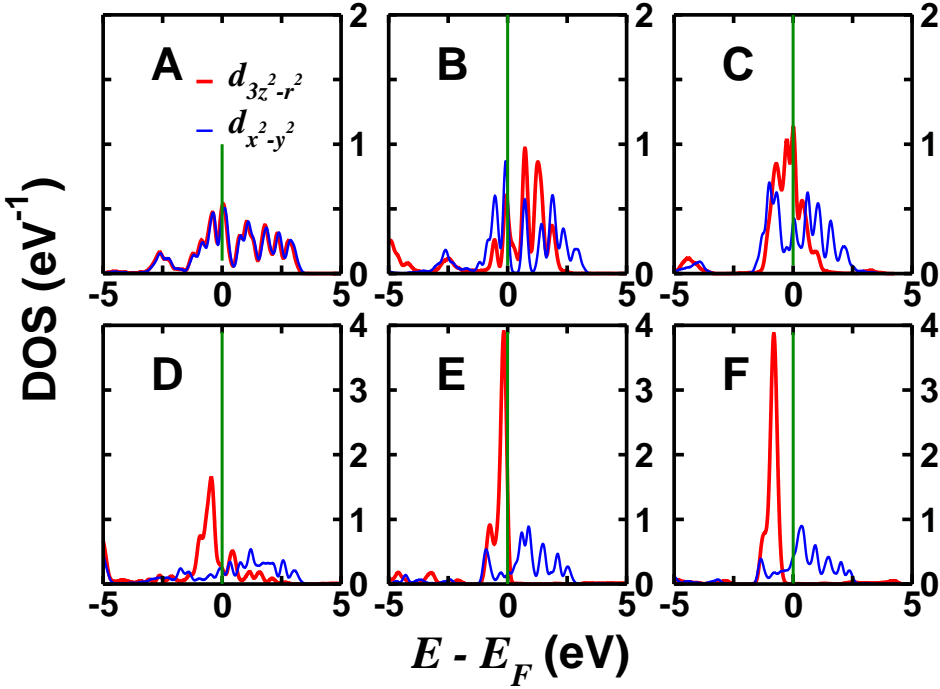}
\caption{\label{fig:dos} Atomic projected density of states: Ni
$d_{3z^2-r^2}$ in red and Ni $d_{x^2-y^2}$ in blue. \textbf{A})
Tetragonal LaNiO$_3$. \textbf{B}) (LaAlO$_3$)$_1$/(LaNiO$_3$)$_1$
superlattice. \textbf{C}) (LaTiO$_3$)$_1$/(LaNiO$_3$)$_1$
superlattice. \textbf{D}) NiO$_2$-terminated LaNiO$_3$ thin film
on a LaAlO$_3$ substrate (only surface Ni is shown). \textbf{E})
(LaTiO$_3$)$_1$/(LaNiO$_3$)$_1$/(RbF)$_2$ superlattice.
\textbf{F}) Ba doped (LaTiO$_3$)$_1$/(LaNiO$_3$)$_1$/(RbF)$_2$
superlattice. Vertical solid green lines mark the Fermi level (set
to zero energy).}
\end{figure}

We begin with bulk LaNiO$_3$ strained to the LaAlO$_3$ substrate,
which makes the LaNiO$_3$ weakly tetragonal ($c/a$=1.01), since
the strain mismatch is small. 
Unstrained bulk LaNiO$_3$ has space group $R\bar{3}c$. 
Table~\ref{tab:all} shows a
small difference between in-plane and out-of-plane Ni-O bond
lengths. Fig.~\ref{fig:dos}A shows the $e_g$-projected electronic
density of states (DOS); the two $e_g$ DOS are
similar, signaling negligible orbital splitting. The Wannier
functions quantify this difference: Table~\ref{tab:all} shows a
0.07 eV splitting of the onsite $e_g$ energies, which is to be
compared to the $e_g$ bandwidth of $\sim$4 eV.

\begin{figure}[t]
\includegraphics[angle=-90,width=16cm]{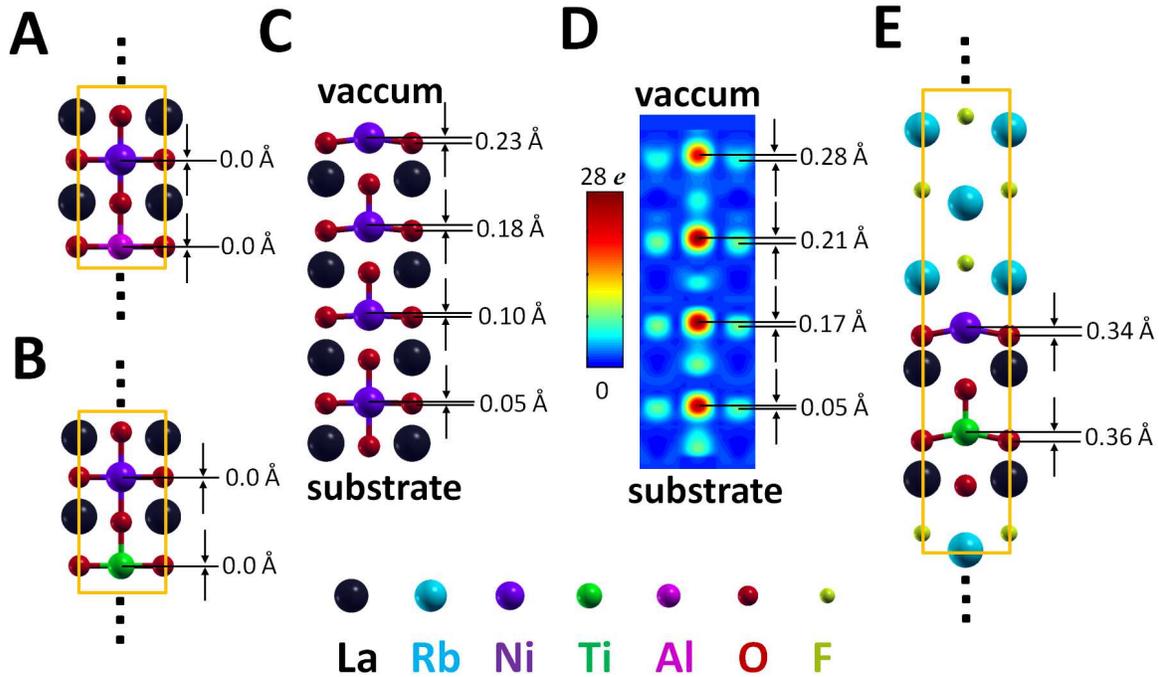}
\caption{\label{fig:structure} Theoretically relaxed structures
\textbf{A}-\textbf{C} and \textbf{E}, as well as the
experimentally determined thin-film structure \textbf{D} by
synchroton x-ray diffraction: \textbf{A})
(LaAlO$_3$)$_1$/(LaNiO$_3$)$_1$ superlattice. \textbf{B})
(LaTiO$_3$)$_1$/(LaNiO$_3$)$_1$ superlattice. \textbf{C})
NiO$_2$-terminated 4 unit cell LaNiO$_3$ thin film on a LaAlO$_3$
substrate (substrate not shown). \textbf{D}) Same as \textbf{C}.
The electron density map shown is a slice through the (101) plane
of the LaNiO$_3$ thin film. \textbf{E})
(LaTiO$_3$)$_1$/(LaNiO$_3$)$_1$/(RbF)$_2$ superlattice.}
\end{figure}

Moving to the LaNiO$_3$/LaAlO$_3$ superlattice
(Fig.~\ref{fig:structure}A), we find reduced out-of-plane hopping,
as evidenced by the narrowed DOS of the $d_{3z^2-r^2}$ band
(Fig.~\ref{fig:dos}B).  However, the Ni is bulk-like,
with uniform Ni-O bond lengths and
nearly degenerate $e_g$ orbitals (Table~\ref{tab:all}). There
is a wider $d_{x^2-y^2}$ band and narrower $d_{3z^2-r^2}$ band,
but the Fermi level cuts through the center of both bands, 
which is consistent with previous density functional theory calculations
~\cite{Hansmann-PRL-2009, han_chemical_2010}.

Next, we consider an LaNiO$_3$/LaTiO$_3$ superlattice
(Fig.~\ref{fig:structure}B). Bulk LaTiO$_3$ has space group $Pbnm$. 
Examination of the Ti $d$
DOS~\cite{supplement} places them above the Fermi energy, thereby
showing successful donation of the electron from Ti to Ni, as
planned (see~\cite{bi-color} for the effect of electron
correlations on the electron transfer). However, the Ni DOS 
(Fig.~\ref{fig:dos}C) and
the relatively small energy splitting (Table~\ref{tab:all}) show
that although the $d_{3z^2-r^2}$ band is filled more than before,
both $e_g$ bands still contribute at the Fermi level. The doping
is successful but insufficient to remove orbital degeneracy.

All of the above structures have inversion symmetry.  A
stoichiometric LaNiO$_3$ film breaks such symmetry with
significant consequences. Fig.~\ref{fig:structure}C shows the
structure of an epitaxial four unit cell NiO$_2$-terminated
LaNiO$_3$ thin film on a LaAlO$_3$ substrate (modeled as a six
unit cell slab of LaAlO$_3$). A polar distortion near the surface
is clearly visible in Fig.~\ref{fig:structure}C that is due to the
polar nature of stoichiometric LaNiO$_3$ films along the (001)
direction~\cite{screening}, with alternating (LaO)$^+$ and
(NiO$_2$)$^-$ atomic planes, which creates a polar electric field
pointing to the surface. As seen in Table~\ref{tab:all}, this
polar field causes a structural distortion that elongates the Ni-O
bond of the surface Ni atom with the O below it. However, more
important than the polarity is the fact that the surface Ni is
missing an oxygen nearest neighbor and thus a Ni-O bond.
Fig.~\ref{fig:dos}D shows a DOS with a narrow $d_{3z^2-r^2}$ band
mostly below the Fermi energy and a $d_{x^2-y^2}$ band mostly
above due to a large orbital energy splitting of 1.29 eV
 (Table~\ref{tab:all}). The reason for the large splitting
is two-fold. Foremost is that the formation of the surface has
eliminated the apical Ni-O bond: the elimination of the Ni
$d_{3z^2-r^2}$-O $p_z$ hopping element lowers the $d_{3z^2-r^2}$
energy (since Ni $e_g$ states are anti-bonding in nature). A
secondary effect is the elongation of the Ni-O bond of the Ni with
the O atom below:  this elongation reduces the same hopping
element and further lowers the $d_{3z^2-r^2}$ energy.  This
surface effect, whereby broken or elongated Ni-O bonds create a
large orbital polarization, is the key that opens the door to the
engineered tri-component superlattices.

Because of the critical importance of this thin film structure to
our approach, we have experimentally grown and characterized this
system, as shown in Fig.~\ref{fig:structure}D. The LaNiO$_3$ film
is NiO$_2$ terminated: the polar distortions (Ni-O $z$
separations) in each NiO$_2$ layer compare well to the theory, and
in both cases the polar distortions decay within a few unit cells
of the surface. These data demonstrate that such thin film
structures are physically realizable and have the atomic-scale
structures that theory predicts.  This comparison and benchmarking
of the theory is important since we will be considering non-bulk structures
and electronic configurations (e.g.,  Ni close to a formal +2 state and Ti 
close to a formal +4 state due to the electron transfer).

Based on all of this information, we now describe the
tri-component superlattices (Fig.~\ref{fig:schematic}D), where we
replace the vacuum in the thin film system by a wide-gap
insulator. We have tested five candidates: LaAlO$_3$, SrTiO$_3$,
BaO, RbF, and NaCl, which all have a good lattice match to
LaAlO$_3$. All five show the same qualitative behavior described
below. We focus on RbF, which has a very large
gap~\cite{laoinsulator}. Fig.~\ref{fig:structure}E shows a relaxed
(LaTiO$_3$)$_1$/(LaNiO$_3$)$_1$/(RbF)$_2$ superlattice. The
structure shows significant polar displacements, indicating
internal electric fields. The displacements are consistent with
the electric field pattern in Fig.~\ref{fig:schematic}D, since the
LaTiO$_3$ and LaNiO$_3$ layers have the La, Ti, and Ni displaced
above the O (upwards electric field), while in the RbF the Rb
atoms are displaced below the F atoms. The alternating direction
of the electric field is due to the doping effect combined with
the periodicity of the superlattice geometry: the
Ti$\rightarrow$Ni electron transfer creates a net field pointing
from the LaTiO$_3$ to LaNiO$_3$ regions, and the periodicity
requires an opposite field in the insulator. The bond lengths of
the Ni with its neighboring anions show a large asymmetry between
in-plane and out-of-plane directions, leading to a significant
orbital splitting of 1.25 eV (Table~\ref{tab:all}).
Fig~\ref{fig:dos}E shows the DOS for this tri-component
superlattice: the narrow $d_{3z^2-r^2}$ band is essentially filled
in the background of a wide $d_{x^2-y^2}$ band. 
Further electron doping can yield a single band system at the Fermi
level with full orbital polarization. To accomplish that, we dope the
RbF with Ba at an areal density of 0.25 Ba per Ni which requires a
much larger in-plane unit cell (see~\cite{supplement} for details), 
with the resulting DOS shown in Fig.~\ref{fig:dos}F.
The doping rigidly
shifts the Fermi energy and fills the $d_{3z^2-r^2}$ band. The
Fermi level now cuts only through a single band, the $d_{x^2-y^2}$
band, in direct analogy to cuprates.

\begin{table}[t]
\caption{\label{tab:all} Onsite energy differences $\Delta$
(second column), and Ni-anion bond lengths $l$ (third and fourth
columns) for the structures studied. The energy difference $\Delta
= E(d_{x^2-y^2}) - E(d_{3z^2-r^2})$ is between the two Ni $e_g$
maximally localized Wannier functions that describe the
anti-bonding conduction bands. For the tri-component superlattice
(last row), the out-of-plane bonds are Ni-O and Ni-F in that
order.}
\begin{center}
\begin{tabular}{c|c|c|c}
\hline
\hline
System                                   & $\Delta$ (eV) & in-plane $l$ (\AA) & out-of-plane $l$ (\AA)  \\
\hline
Tetragonal LaNiO$_3$                     &      0.07      & 1.86  & 1.88       \\
\hline
(LaAlO$_3$)$_1$/(LaNiO$_3$)$_1$          &      0.08      & 1.86  & 1.86      \\
\hline
(LaTiO$_3$)$_1$/(LaNiO$_3$)$_1$          &      0.23      & 1.86  & 2.00    \\
\hline
NiO$_2$-terminated LaNiO$_3$ thin film   &      1.29      & 1.87  & 2.01     \\
\hline
(LaTiO$_3$)$_1$/(LaNiO$_3$)$_1$/(RbF)$_2$ &     1.25      & 1.89  & 2.64/2.76  \\
\hline
\hline
\end{tabular}
\end{center}
\end{table}

We remark that while experimental realization of these
tri-component superlattices will be challenging due to the
complexity of their structure, our  approach is general and
flexible, which permits the consideration of many possible
materials combinations.  For example, broken inversion symmetry
occurs in the superconductors Li$_2$Pd$_3$B and Li$_2$Pt$_3$B
where the lower symmetry causes Copper pairs to display interesting electronic
 properties~\cite{Yuan-PRL-2006}.

In summary, we describe a general approach to realizing the
single-band 2D Hubbard model on nickelate conducting planes.  The key
is to engineer the atomic-scale structure around the Ni to elongate
some Ni-anion bonds compared to others. This approach creates a large
$e_g$ orbital energy splitting and orbital polarization, much like
what is found in cuprate high-temperature superconductors.  The two
main tools are broadly applicable and robust: (i) charge transfer via
doping (resulting in electric fields and polar displacements), and
(ii) use of a tri-component superlattice to break inversion
symmetry. Given the generality of the approach, it can be applied to
other systems to similarly engineer orbital polarizations.

\begin{acknowledgments}
Work at Yale is supported by grant W911NF-10-1-0206 from the Army Research Office, with funding from the DARPA OLE Program, and NSF DMR 1119826 (CRISP). Computational facilities are
supported by NSF grant CNS 08-21132 and by the facilities and staff of the Yale
University Faculty of Arts and Sciences High Performance Computing
Center.  Additional computations are carried out via the NSF
TeraGrid and XSEDE resources through grant TG-MCA08X007.  Use of
the Advanced Photon Source, an Office of Science User Facility
operated for the U.S. Department of Energy (DOE) Office of Science
by Argonne National Laboratory, is supported by the U.S. DOE 
grant DE-AC02-06CH11357.
\end{acknowledgments}


\begin{thebibliography}{50}
\bibitem[{str()}]{strain}
\bibinfo{note}{Our key results do not rely on strain engineering and are
  expected to be robust versus epitaxial strain.}
\bibitem[{scr()}]{screening}
\bibinfo{note}{Bulk LaNiO$_3$ is metallic but has a finite screening length.
  Thus within a screening length of the polar (001) surface, one finds a
  non-negligible polar field which produces polar distortions of the ionic
  positions.}
\bibitem[{lao()}]{laoinsulator}
\bibinfo{note}{Experimentally, we expect LaAlO$_3$ to be a logical starting
  choice for insulator since the superlattice is then composed of the most
  structurally similar materials.}
\end{thebibliography}

\end{document}